\documentclass[aps,prd,showpacs,nofootinbib]{revtex4}

\usepackage{amssymb}
\usepackage{amsmath}
\usepackage[dvipsone]{graphics}
\usepackage{psfig}
\usepackage{epsfig}
\usepackage{bm}
\expandafter\ifx\csname package@font\endcsname\relax\else
\expandafter\expandafter
 \expandafter\usepackage
 \expandafter\expandafter
 \expandafter{\csname package@font\endcsname}%
\fi
\begin{document}
\title{Laser Ranging Delay in the Bi-Metric Theory of Gravity}
\author{Sergei M. Kopeikin}
\email{kopeikins@missouri.edu}
\affiliation{Department of Physics \& Astronomy, University of
Missouri-Columbia, Columbia, MO 65211, USA}
\author{Wei-Tou Ni}
\email{wtni@pmo.ac.cn}
\affiliation{Purple Mountain Observatory, Chinese Academy of Sciences, Nanjing, 210008, China }
\date{\today}
\begin{abstract}
We introduce a linearized bi-metric theory of gravity with two metrics. The metric $g_{\alpha\beta}$ describes null hypersurfaces of the gravitational field while light moves on null hypersurfaces of the {\it optical} metric $\bar{g}_{\alpha\beta}$. Bi-metrism naturally arises in vector-tensor theories with matter being non-minimally coupled to gravity via long-range vector field. We derive explicit Lorentz-invariant solution for a light ray propagating in space-time of the bi-metric theory and disentangle relativistic effects associated with the existence of the two metrics. This anlysis may be valuable for future spaceborne laser missions ASTROD and LATOR dedicated to map various relativistic gravity parameters in the solar system to unparalleled degree of accuracy.  
\end{abstract}
\maketitle
Recently Carlip \cite{carlip} has introduced a bi-metric theory of gravity with two metrics, $g_{\alpha\beta}$ and $\bar{g}_{\alpha\beta}$, and a unit vector field $w^\alpha$ coupled to matter via constant parameter $\epsilon$. This theory is a variant of a vector-tensor theory \cite{jack} where the vector field $w^\alpha$ obeys the source-free field equations and is responsible for the spontaneous violation of the Lorentz invariance of gravity \cite{kost,lam} in the sense that it introduces a preferred frame which effects can be observed only in gravitational experiments conducted in non-negligible gravitational field. One can show that in the case of a bi-metric theory of gravity the Huges-Drever experiments and other precision experiments constrain the asymptotic (vanishing gravity) difference between these two metrics severely \cite{ni,ni-1,mattingly}. Here we consider more complicated (non-vanishing gravity) case of geometric optics of light rays (laser beams) in the bi-metric theory. Carlip's bi-metric theory \cite{carlip} adopts specific values of parameters in Jackobson's theory \cite{jack} and adds one more parameter $\epsilon$ which is a coupling constant between the vector field $w^\alpha$ and the stress-energy tensor of matter. It is useful to understand what kind of relativistic effects one can expect in the bi-metric theory in application to the spaceborne laser ranging experiments like ASTROD \cite{astrod-1,astrod-2} and LATOR \cite{lator-1,lator-2,lator-3}. We shall investigate this problem by calculating the time delay of light propagating in the time-dependent gravitational field of an arbitrary moving body. 

In what follows, we shall consider an isolated N-body system (a solar system) resided in an asymptotically flat space-time of the gravity metric $g_{\alpha\beta}$. According to Carlip \cite{carlip} the optical metric  
\begin{eqnarray}
\label{p0}
\bar{g}_{\alpha\beta}&=&g_{\alpha\beta}+\left(1-\frac{1}{\epsilon^2}\right)w_\alpha w_\beta\;,\\\label{pik}
\bar{g}^{\alpha\beta}&=&g^{\alpha\beta}-\left(\epsilon^2-1\right)w^\alpha w^\beta\;,
\end{eqnarray}
where $\epsilon$ is a constant parameter defining the coupling of the vector field $w^\alpha$ with matter and describing the degree of violation of the Lorentz invariance for electromagnetic field (and other material fields). Following Carlip \cite{carlip} we assume that the Greek indices are raised and lowered with the metric $g_{\alpha\beta}$. 

Let us work in a global coordinate system $x^\alpha=(ct,x^i)$, where $c$ in Carlip's theory \cite{carlip} is the speed of gravity and $x^i$ are spatial coordinates. In the global frame the linearized expansions for the metric and the vector field are
\begin{eqnarray}
\label{q1}
g_{\alpha\beta}&=&\eta_{\alpha\beta}+h_{\alpha\beta}\;,\\\label{q2}
g^{\alpha\beta}&=&\eta^{\alpha\beta}-h^{\alpha\beta}\;,\\\label{q3}
w^\alpha&=&V^\alpha+\zeta^\alpha\;,\\\label{q4}
w_\alpha&=&V_\alpha+\zeta_\alpha-h_{\alpha\beta}V^\beta\;,
\end{eqnarray}
where $\eta_{\alpha\beta}={\rm diag}(-1,+1,+1,+1)$ is the Minkowski metric, $h_{\alpha\beta}$ is the perturbation of the gravity metric, and $\zeta^\alpha$ is the perturbation of the vector field which unperturbed value in the global frame is $V^\alpha$. We emphasize that $V^\alpha$ remains arbitrary and our analysis is not limited to the case of the preferred frame where $V^\alpha=(1,0,0,0)$. 

The optical metric $\bar{g}_{\alpha\beta}$ is decomposed as follows
\begin{eqnarray}
\label{q5}
\bar{g}_{\alpha\beta}&=&\bar{\eta}_{\alpha\beta}+\bar{h}_{\alpha\beta}\;,\\\label{q6}
\bar{g}^{\alpha\beta}&=&\bar{\eta}^{\alpha\beta}-\bar{h}^{\alpha\beta}\;,
\end{eqnarray}
where the unperturbed part of the optical metric is defined by
\begin{eqnarray}
\label{A7}
\bar{\eta}^{\alpha\beta}&=&{\eta}^{\alpha\beta}-\left(\epsilon^2-1\right)V^\alpha V^\beta\;,\\
\label{A8}
\bar{\eta}_{\alpha\beta}&=&{\eta}_{\alpha\beta}+\left(1-\frac{1}{\epsilon^2}\right)V_\alpha V_\beta\;,
\end{eqnarray}
and the perturbation
\begin{equation}
\label{q7}
\bar{h}^{\alpha\beta}=h^{\alpha\beta}+\left(\epsilon^2-1\right)(V^\alpha \zeta^\beta+V^\beta\zeta^\alpha)\;.
\end{equation}
 
According to Carlip \cite{carlip} light propagates in the bi-metric theory along light geodesics of the optical metric $\bar{g}_{\alpha\beta}$. In geometric optics limit the light rays are defined by a covariant equation for electromagnetic phase (eikonal) $\varphi$ which reads \cite{carlip,synge}
\begin{equation}
\label{A3}
\bar{g}^{\mu\nu}\partial_\mu\varphi\partial_\nu\varphi=0\;.
\end{equation}
This equation is formally equivalent to the equation of light propagating through dispersive medium with refraction index $\epsilon$ moving with respect to a preferred frame with velocity $w^\alpha$. Theory of light propagation through the dispersive medium has been worked out by Synge \cite{synge} and we shall use his theory in order to integrate equation (\ref{A3} and to interpret its solution.  

As follows from equation (\ref{A3}) we do not need to know solutions for the metric perturbation $h^{\alpha\beta}$ and that of the vector field $\zeta^\alpha$ separately. What we need to perform the integration of equation (\ref{A3}), is solution for the perturbation $\bar{h}^{\alpha\beta}$ of the optical metric (\ref{q6}). In the linearized approximation the metric perturbation $\bar{h}^{\alpha\beta}$ obeys the following gravity field equations \cite{carlip}
\begin{eqnarray}
\label{p1}
\Box \bar{h}^{\alpha\beta}&=&-\frac{16\pi G}{(1-4\varpi)\epsilon c^2}\left(S^{(\alpha}_{\;\;\mu}T^{\beta)\mu}-\eta^{\alpha\beta}T^\lambda_{\;\,\lambda}\right)\;,
\end{eqnarray}
where $\Box\equiv -c^{-2}\partial^2/\partial t^2 +\nabla^2$ is the wave
operator in flat space-time defining null characteristics of the gravitational field, the constant tensor $S^{\alpha\beta}=\eta^{\alpha\beta}+2\varpi V^\alpha V^\beta$, $\varpi$ is another constant parameter of the bi-metric theory ($\varpi=1$ in general relativity), and $T^{\mu\nu}$ is the stress-energy
tensor of the point-like bodies composing of the N-body system. Equation (\ref{p1}) is valid under imposing the following gauge conditions 
\begin{equation}
\label{gc}
\partial_\beta\left(h^{\alpha\beta}-\frac{1}{2}\eta^{\alpha\beta}h\right)+\left(1-\frac{1}{\epsilon^2}\right)V_\alpha\partial_\beta\zeta^\beta=0\;.
\end{equation}
In the linearized approximation the stress-energy tensor reads \cite{carlip}
\begin{eqnarray}
\label{hh} 
T^{\alpha\beta}(t, {\bm x})=\sum_{a=1}^N \frac{M_a u_a^\alpha\,
u_a^\beta\, \delta^{(3)}\bigl({\bm x}-{\bm x}_a(t)\bigr)}{\gamma_a\sqrt{1-\left(1-\epsilon^{-2}\right)\left(u^\mu V_\mu\right)^2}}\;,
\end{eqnarray}
where the index $a=1,2,...,N$ enumerates gravitating bodies of the
solar system, $M_a$ is the (constant) rest mass of the $a$th body,
${\bm x}_a(t)$ is time-dependent spatial coordinate of the $a$th body,
${\bm v}_a(t)= d{\bm x}_a(t)/dt$ is velocity of the $a$th body,
$u_a^\alpha=\gamma_a (1,\,{\bm v}_a/c)$ is the four-velocity of the
$a$th body, $\gamma_a=\bigl(1-v_a^2/c^2\bigr)^{-1/2}$ is the
Lorentz-factor, and $\delta^{(3)}({\bm x})$ is the 3-dimensional
Dirac's delta-function.  

Because the field equations (\ref{p1}) are linear, we can consider
their solution as a linear superposition of the solutions for each
body. It allows us to focus on the relativistic effects caused by one
body (Sun, planet, etc.) only.
Solving equations (\ref{p1}) by making use of the retarded Li\'enard-Wiechert tensor potentials \cite{LL}, one obtains the metric
tensor perturbation
\begin{eqnarray}\label{1} 
\bar{h}^{\alpha\beta}(t,{\bm x})&=&\frac{2GM}{(1-4\varpi)c^2}\frac{2u^\alpha u^\beta+\eta^{\alpha\beta}+2\varpi \left(u^{\alpha}V^{\beta}+u^{\beta}V^{\alpha}\right)\left(u^\mu V_\mu\right)}{\sqrt{\epsilon^2-\left(\epsilon^{2}-1\right)\left(u^\mu V_\mu\right)^2}}\frac{1}{r_R}\;,
\end{eqnarray}
where $r_R\equiv -u_\alpha r^\alpha$,
$r^\alpha=x^\alpha-z^\alpha(s)$, $z^\alpha(t)=(ct, {\bm z}(t))$ is the world line of the light-ray deflecting body parametrized by the coordinate time $t$, $u^\alpha(t)=c^{-1}dz^\alpha(t)/dt$. 

Because we solved the field equations
(\ref{p1}) in terms of the retarded Li\'enard-Wiechert potentials, the distance $r^\alpha=x^\alpha-z^\alpha(s)$, the
body's worldline $z^\alpha(s)=(cs, {\bm z}(s))$, and the body's
four-velocity $u^\alpha(s)$ in the equation (\ref{1}) are functions of the {\it retarded} time
$s$. The retarded time $s$ is found in the linearized approximation of the bi-metric theory as a solution of the gravity null cone equation
\begin{equation}
\label{grav}
\eta_{\mu\nu}r^\mu
r^\nu\equiv\eta_{\mu\nu}\Bigl(x^\mu-z^\mu(s)\Bigr)\Bigl(x^\nu-z^\nu(s)\Bigr)=0\;,
\end{equation}
that is
\begin{equation}
\label{1a}
s=t-\frac{1}{c}|{\bm x}-{\bm z}(s)|\;,
\end{equation}
where the fundamental constant $c$ in equation (\ref{1a}) is
the fundamantal speed of propagation of gravity. 

\begin{figure*}
\includegraphics[keepaspectratio=true,width=15cm,height=15cm]{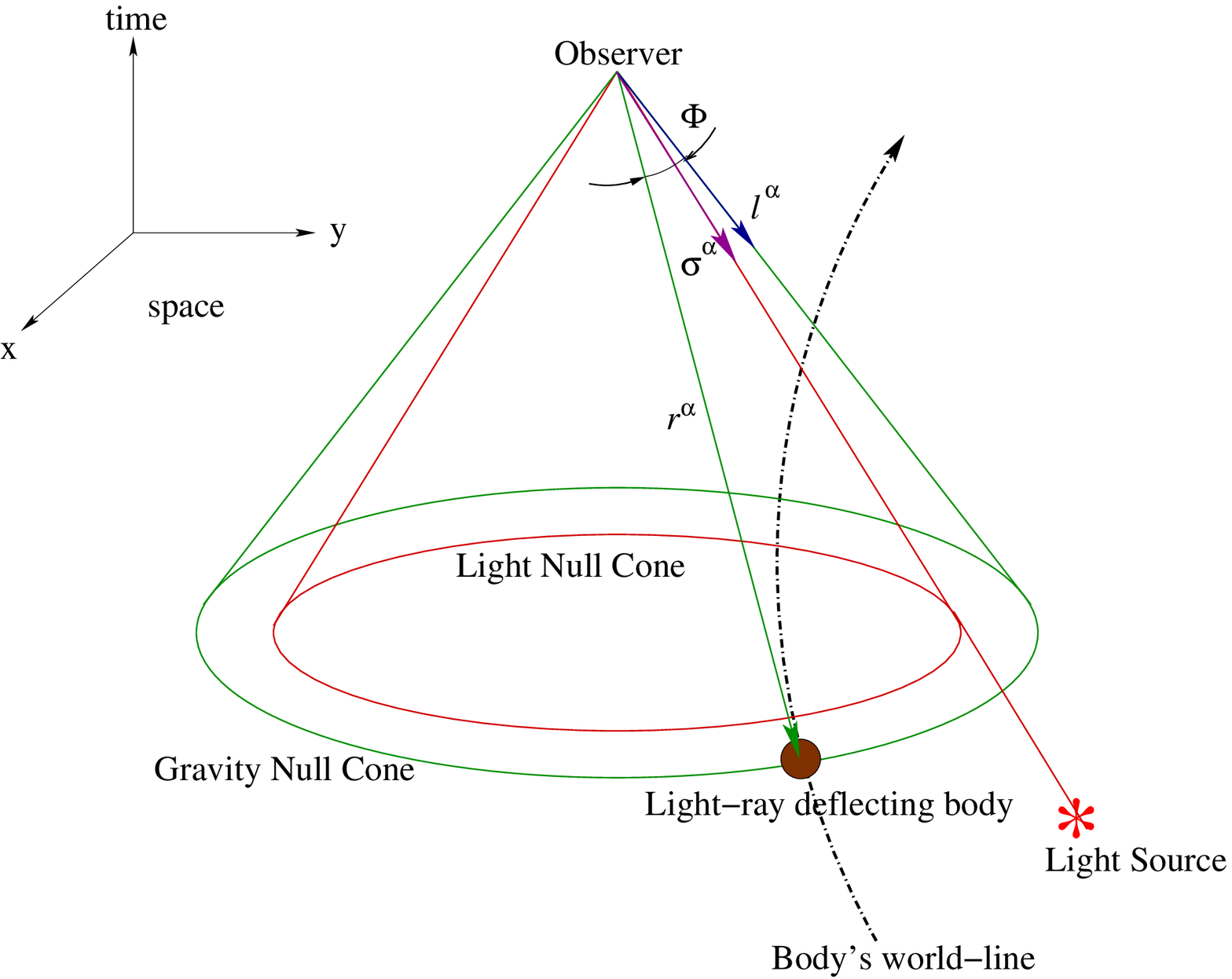}\vspace{1cm}
\caption{\label{twocones}
The light and gravity null cones of the bi-metric theory are shown. Grvity propagates from the massive body to observer along the gravity null cone defined by the metric ${g}_{\alpha\beta}$. When gravity is switched off, the unperturbed direction to the source of light is defined by vector $\sigma^\alpha$ lying on the null hypersurface of the optical metric $\bar{g}_{\alpha\beta}$. Gravity perturbs the direction of the light propagation and changes it from $\sigma^\alpha$ to $l^\alpha$ which belongs to the gravity null cone of the metric ${g}_{\alpha\beta}$. Moving light-ray deflecting body (Sun, planet) deflects light from its retarded position defined with respect to observer by a null vector $r^\alpha=x^\alpha-x^\alpha_J(s)$ which resides in the gravity null cone. Gravitational perturbation of the light eikonal is $\psi=-(2GM/c^2)\chi\ln\Phi$, where $\Phi=-l_\alpha r^\alpha$. Light-ray deflection experiments measure the range $\chi$ and shape $\Phi$ of the gravitational time delay of light. The range measurement allows us to pin down the overall magnitude of the delay while its shape $\Phi$ measures direction of the vector $l^\alpha$ under assumption that vector $r^\alpha$ is known from the solar system ephemeris. Divergence of $l^\alpha$ from $\sigma^\alpha$ is proportional to the degree of violation of the Lorentz invariance in Einstein's gravity theory with respect to Maxwell's electrodynamics.}
\end{figure*}

Light rays are defined by a covariant equation (\ref{A3}) for electromagnetic phase (eikonal) $\varphi$. Assuming that unperturbed solution of equation (\ref{A3}) is a plane wave we can write a general solution of this equations as follows
\begin{equation}
\label{A4}
\varphi(x^\alpha)=\varphi_0+k_\alpha x^\alpha+\psi(x^\alpha)\;,
\end{equation}
where $k_\alpha$ is an unperturbed (constant) wave co-vector of the electromagentic wave, and $\psi(x)$ is a relativistic perturbation of the eikonal generated by the metric tensor perturbation $\bar{h}_{\alpha\beta}$ defined in equation (\ref{1}). Substitution of equation (\ref{A4}) to (\ref{A3}) yields
\begin{eqnarray}\label{A5}
\bar{\eta}^{\alpha\beta}k_\alpha k_\beta&=&0\;,\\
\label{A6}
\bar{\eta}^{\alpha\beta}k_\alpha\frac{\partial\psi}{\partial x^\beta}&=&\frac{1}{2}{h}^{\alpha\beta}k_\alpha k_\beta\;.
\end{eqnarray} 
Let us define a vector (see Fig. \ref{twocones})
\begin{equation}
\label{A9}
\sigma^\alpha=\bar{\eta}^{\alpha\beta}k_\beta=k^\alpha-\left(\epsilon^2-1\right)\left(V^\beta k_\beta\right)V^\alpha\;,
\end{equation}
such that 
\begin{equation}\label{A10}
\bar\eta_{\alpha\beta}\sigma^\alpha\sigma^\beta=0\;,\qquad\qquad\mbox{and}\qquad\qquad k_\alpha\sigma^\alpha=0\;.
\end{equation}
Vector $\sigma^\alpha$ defines the direction of propagation of light ray from a source of light (laser, star) to observer (see Fig. \ref{twocones}). Making use of vector $\sigma^\alpha$ simplifies equation (\ref{A6}) and reduces it to the following form
\begin{equation}\label{A11}
\sigma^\alpha\frac{\partial\psi}{\partial x^\alpha}=\frac{1}{2}{h}_{\alpha\beta}\sigma^\alpha \sigma^\beta\;.
\end{equation}
Unperturbed characteristics of the eikonal equation (\ref{A11}) are straight lines (light rays) parametrized by the affine parameter $\lambda$ in such a way that 
\begin{equation}\label{A12}
\frac{d}{d\lambda}=\sigma^\alpha\frac{\partial}{\partial x^\alpha}\;.
\end{equation}
Integration of the equation (\ref{A12}) by making use of the unperturbed characteristics is straightforward (see, for example, \cite{zm}) and 
can be written as follows
\begin{equation}\label{Api}
\psi(x^\alpha)=-\frac{2GM}{c^2}\chi  \ln\Bigl(-{l}_\alpha r^\alpha \Bigr)\;,
\end{equation}
where 
\begin{eqnarray}\label{iop}
\chi&=&\frac{\left(\sigma_\alpha u^\alpha\right)^2+(1/2)\left(\sigma_\alpha \sigma^\alpha\right) +2\varpi\left(\sigma_\alpha u^\alpha\right)\left(\sigma_\alpha V^\alpha\right)\left(u_\alpha V^\alpha\right)}{(1-4\varpi)(\epsilon^2-\left(\epsilon^{-2}-1\right)\left(u^\mu\zeta_\mu\right)^2)^{1/2}\left(P_{\alpha\beta}\sigma^\alpha\sigma^\beta\right)^{1/2}}\;,\\
\label{A14}
{l}^\alpha&=&\sigma_\bot^\alpha+\sigma_\bot u^\alpha\;,\\
\label{A15}
\sigma_\bot^\alpha&=&P^{\alpha}_{\;\beta}\sigma^\beta\;,\\
\label{A16}
\sigma_\bot&=&\left(\sigma_{\bot\alpha}\sigma^\alpha_\bot\right)^{1/2}=\left(P_{\alpha\beta}\sigma^\alpha\sigma^\beta\right)^{1/2}\;,
\end{eqnarray}
and 
\begin{equation}\label{A17}
P_{\alpha\beta}=\eta_{\alpha\beta}+u_\alpha u_\beta\;,
\end{equation}
is the operator of projection on the hyperplane orthogonal to the four-velocity $u^\alpha$ of the light-ray deflecting body ($P_{\alpha\beta}P^\beta_{\;\gamma}=P_{\alpha\gamma}$). It is easy to confirm that solution (\ref{Api}) is valid by observing that 
\begin{equation}
\label{gimk}
\frac{d}{d\lambda}\ln\Bigl(-l_\alpha r^\alpha\Bigr)=-\frac{\sigma_\bot}{r_R}\;,
\end{equation}
where $r_R=-u_\alpha r^\alpha$, and equations 
\begin{eqnarray}
\label{qa}
\partial_\alpha r^\mu&=&\delta^\mu_\alpha-\frac{u^\mu}{\gamma}\partial_\alpha s\;,\\ 
\label{po} 
\partial_\alpha s&=&-\gamma\,\frac{r_\alpha}{r_R}\;,
\end{eqnarray}
where $\gamma=(1-\beta^2)^{-1/2}$, have been used.
Equation (\ref{A14}) allows us to recast the argument of the logarithm in equation (\ref{Api}) as
\begin{equation}\label{A18}
{l}_\alpha r^\alpha=\sigma_\bot^\alpha r_\alpha-\sigma_\bot r_R\;.
\end{equation}
It is remarkable that both vectors ${l}^\alpha$ and $r^\alpha$ are null vectors of the gravity metric $g_{\alpha\beta}$ (see Fig. \ref{twocones}). Indeed, in the linearized approximation $g_{\alpha\beta}=\eta_{\alpha\beta}$ and one can easily prove by inspection that 
\begin{eqnarray}\label{A19}
\eta_{\alpha\beta}{l}^\alpha{l}^\beta&=&0\;,\\
\label{A20}
\eta_{\alpha\beta}r^\alpha r^\beta&=&0\;,
\end{eqnarray}
which are consequences of the definitions given by equations (\ref{grav}) and (\ref{A14}). Thus, neither $l^\alpha$ nor $r^\alpha$ belong to the null cone of the optical metric $\bar{g}_{\alpha\beta}$ but characterize the null hypersurfaces of the gravity metric $g_{\alpha\beta}$. 

Solution (\ref{A4}), (\ref{Api}) for the electromagnetic eikonal in the bi-metric theory should be compared with a similar solution for the case of propagation of light in general relativity where the gravity and light null cones coincide \cite{ks}. The reader can see that the null characteristics of the gravity metric $g_{\alpha\beta}$  enters the gravitationally perturbed part of the eikonal (\ref{Api}) in the bi-metric theory in the form of the dot product ${l}_\alpha r^\alpha$ which is the argument of the logarithm, where $r^\alpha$ is the null distance of the metric $g_{\alpha\beta}$ between the observer and the light-ray deflecting body. A remarkable fact is that both $l^\alpha$ and $r^\alpha$ are null vectors of the metric $g_{\alpha\beta}$ describing null hypersurfaces of the gravitational field. Consequently, gravitational light-ray deflection experiments in the field of moving bodies are sensitive to, and can measure, the divergence between the null characteristics of the gravity metric $g_{\alpha\beta}$ and the optical metric $\bar{g}_{\alpha\beta}$ in the case of a non-stationary gravity field in contrast to other relativistic experiments limiting the PPN preferred frame parameters $\alpha_1,\alpha_2,\alpha_3$ \cite{mattingly}. This allows us to measure the spontaneous violation of the Lorentz invariance of the gravitational field predicted by the vector-tensor theories of gravity that admit existence of a vector field $w^\alpha$ coupled to matter via parameter $\epsilon$ parametrizing the difference between the gravity and optical metrics. In conventional type of the gravitational light-ray deflection experiments conducted with VLBI \cite{ks,km} one needs the angle $\Phi$ (see Fig. \ref{twocones}) to be as small as possible to magnify the Lorentz-invariance violation effects driven by gravity field. Currently, VLBI can measure gravitomagnetic effects of order $v/c$ beyond the static Shapiro effect \cite{fk,kop-ijmp}. Further progress in measuring more subtle effects of the bi-metric theory of gravity of order $v^2/c^2$ and higher beyond the static Shapiro time delay can be achieved with laser ranging technique in the experiments like LATOR \cite{lator-1,lator-2,lator-3} and/or ASTROD \cite{astrod-1,astrod-2}. 
In case of laser ranging between spacecrafts with a gravitating body (Sun) located near the direction of the laser beam, the angle $\Phi$ (see Fig. \ref{twocones}) can vary in a large dynamical range so that relativistic effects of the bi-metric theory of gravity could be explored with much better precision than in VLBI experiments. We will study this situation for laser ranging experiment in various theories, e.g., in the more general vector-metric theories \cite{lxh} and in the axion electrodynamics \cite{ni-2}.

\newpage
\end{document}